\documentclass{aa}
\usepackage{rotating}
\usepackage{natbib}
\bibpunct{(}{)}{;}{a}{}{,}
\usepackage{txfonts}
\usepackage{graphicx}

\begin{document}

\authorrunning{de Jong et al.}

\title{Ground-based variability surveys towards Centaurus A:
  worthwhile or not?}

\author{Jelte~T.A.~de~Jong\inst{1,2}
        \and Konrad~H.~Kuijken\inst{3,2}
        \and Philippe~H\'eraudeau\inst{4,2}
}

\offprints{Jelte~T.~A.~de~Jong, \email{dejong@mpia.de}}

\institute{Max-Planck-Institut f\"ur Astronomie, K\"onigstuhl 17,
  D-69117 Heidelberg, Germany
\and
Kapteyn Astronomical Institute, University of Groningen, PO Box 800,
  9700 AV, Groningen, the Netherlands
\and
Sterrewacht Leiden, University of Leiden, PO Box 9513, 2300 RA, Leiden,
the Netherlands
\and
Argelander Institut f\"ur Astronomie, Auf dem H\"ugel 71, D-53121 Bonn, Germany}

\date{Received  / Accepted }

\abstract
{Difference imaging has proven to be a powerful technique for
  detecting and monitoring the variability of unresolved stellar sources
  in M 31. Using this technique in surveys of galaxies outside the
  Local Group could have many interesting applications.}
{The goal of this paper is to test difference imaging photometry on
  Centaurus A, the nearest giant elliptical galaxy, at a distance of 4
  Mpc.}
{We obtained deep photometric data with the Wide
  Field Imager at the ESO/MPG 2.2m at La Silla spread over almost two
  months. Applying the difference imaging photometry package DIFIMPHOT,
  we produced high-quality difference images and detected variable
  sources. The sensitivity of the current observational setup was
  determined through artificial residual tests.}
{In the resulting high-quality difference images, we detect 271
  variable stars. We find a difference flux detection limit
  corresponding to $m_R\simeq$24.5. Based on a simple model of the
  halo of Centaurus A, we estimate that a ground-based microlensing
  survey would detect in the order of 4 microlensing events per year
  due to lenses in the halo.}
{Difference imaging photometry works very well at the distance of
  Centaurus A and promises to be a useful tool for detecting and studying
  variable stars in galaxies outside the local group. For microlensing
  surveys, a higher sensitivity is needed than achieved here, which
  would be possible with a large ground-based telescope or space
  observatory with wide-field imaging capabilities.}


\maketitle

\section{Introduction}
\label{sec:intro}

Using difference imaging techniques
\citep{tomaneycrotts,gould96,alard00} it is possible to detect and
monitor stellar variability in highly crowded fields where the
individual stars are unresolved. During the past decade these
techniques have been used to this aim by several microlensing surveys
towards the \object{Magellanic Clouds} \citep[e.g.][]{alcock00,tisserand07} and
the Andromeda galaxy (\object{M 31})
\citep[e.g.][]{riffeser03,calchi05,dejong06}. This has led to the
successful detection of several microlensing events and tens of
thousands of previously unknown variable stars in M31. Following these
results the question arises whether the application of these
techniques can be extended to even more distant objects. Both the
detection of variable stars and of microlensing in galaxies outside
the local group would have several interesting applications.

The number of known variable stars in galaxies could be multiplied,
which would be interesting in itself, and also useful for, for
example, distance determination and the study of stellar populations
and star formation histories.  Microlensing surveys outside the Local
Group would also be an important tool to study the luminosity
functions and the halo compact object content of galaxies other than
the Milky Way and M 31. On the one hand this would be useful as a
further measure of the contribution of massive compact halo objects
(MACHOs) to galaxy halos.  On the other hand such microlensing surveys
could be an important step to the generalisation of the picture
obtained from the surveys in the Local Group. Both the Milky Way and
M31 are two large spiral galaxies in a relatively low-density
environment. It would be interesting to also study the compact object
content in the halos of galaxies in other environments or in the halos
of other types of galaxies.

If the dark matter does indeed have a baryonic MACHO contribution,
this component could be different in elliptical galaxies compared to
spirals, since they are believed to have undergone different formation
histories and star formation episodes. Furthermore, elliptical
galaxies tend to live in more dense environments, where galactic
encounters are more common and old stellar populations would have been
pushed out to larger radii into the halo. Microlensing surveys in
elliptical galaxies would also enable the study of these stellar
halos, that might be important repositories of faint stellar mass.

A first attempt at a microlensing survey in an elliptical galaxy was
made by \cite{baltz04}. Using observations taken with the Wide Field
and Planetary Camera 2 (WFPC2) on the Hubble Space Telescope, they
performed a search for microlensing events in the giant elliptical
galaxy \object{M 87}.  Due to the large distance to M 87 of about 16 Mpc
\citep{tonry01} doing this experiment proved difficult in practice. In
their study, \cite{baltz04} find only seven variable sources, 1 of which
is consistent with microlensing, although it is bluer than expected
for a typical microlensing event.

In this paper we present a pilot study of a ground-based variability
study using difference imaging photometry towards the giant elliptical
galaxy \object{Centaurus A} (Cen A, \object{NGC 5128}). Apart from the
general interest of a variable star and microlensing study in an
elliptical galaxy discussed above, this survey would have additional
scientific interests. From the warped gas layer, the outer isophotes
and the kinematics of the planetary nebulae, there are indications
that the potential and therefore the halo around Centaurus A is
triaxial \citep{hui95}. This means that the lines-of-sight toward
different parts of the galaxy have different path-lengths through the
halo. Because of this, the microlensing event rate will be asymmetric
over the face of Cen A, much like the microlensing rate due to halo
lensing is asymmetric between the near and far side of the disk of M
31. The spatial distribution of halo microlensing events could
therefore also be used to constrain the shape of the dark halo of Cen
A.  Because the distance to Cen A \citep[4 Mpc,][]{tonry01} is much
smaller than the distance to M 87, the sensitivity will be much higher
and the crowding of variable sources much less strong. Nevertheless,
Cen A lies five times further away than M 31, so the project is not
without challenge.

Section \ref{sec:data} describes the data and methods that were used
in this study. In Section \ref{subsec:variables} the resulting
difference images are described and the detected variable sources
presented. We determine the detection efficiency for variable sources
in Section \ref{subsec:deteff}. The prospects of ground-based (halo)
microlensing surveys towards Centaurus A are estimated in Section
\ref{sec:microlensing}, based on our results and a simple model of the
halo of Centaurus A. Finally, in Section \ref{sec:conclusions} we
present our conclusions.

\section{Data and methods}
\label{sec:data}

Goal of the observational set-up was to obtain deep, 9\,000$s$,
photometry in the broad $R_c$ filter during each of a series of nights
spread over a several month period, resulting in both high sensitivity
and sufficient temporal coverage to detect faint variable sources.
Observations were done during 12 nights spread between April 15th and
June 8th 2005 with the Wide Field Imager \citep[WFI;][]{wfi} at the
ESO/MPG 2.2m telescope at La Silla, Chile.  With its field-of-view of
34\arcmin$\times$33\arcmin (pixel scale 0.238\arcsec per pixel) the
WFI enables the monitoring of Cen A in one single pointing. The total
exposure time per night was divided over 1\,000$s$ exposures to help
with cosmic ray rejection. A small subset of data was discarded
because of poor seeing ($>$1.3\arcsec), leaving data for 10 nights,
and 9 nights with the total exposure time of 8\,000$s$ or longer. A
summary of the exposure times and average seeings is listed in Table
\ref{tab:data}. The seeings listed are the average
of all exposures from the four central chips in the array, with the
quoted errors the standard deviation.

\begin{table}
\begin{center}
\caption{Summary of data used}
\begin{tabular}{lccc}
\hline
\hline
Date & Exposures & Total $t_{exp}$ & Seeing \\
~ & ~ & (sec) & (\arcsec) \\
\hline
April 17 2005 & 4 & 4\,000 & 0.99$\pm$0.06 \\
April 30 2005 & 9 & 9\,000 & 0.83$\pm$0.10 \\
May 9 2005 & 8 & 8\,000 & 0.78$\pm$0.03 \\
May 11 2005 & 9 & 9\,000 & 0.84$\pm$0.13 \\
May 12 2005 & 9 & 9\,000 & 0.92$\pm$0.07 \\
May 29 2005 & 9 & 9\,000 & 1.00$\pm$0.14 \\
May 30 2005 & 9 & 9\,000 & 0.82$\pm$0.10 \\
May 31 2005 & 9 & 9\,000 & 0.80$\pm$0.13 \\
June 1 2005 & 9 & 9\,000 & 1.02$\pm$0.15 \\
June 8 2005 & 9 & 9\,000 & 0.79$\pm$0.07 \\
\hline
\end{tabular}
\label{tab:data}
\end{center}
\end{table}

The data reduction was performed with the Astronomical Wide-Field
Imaging System for Europe\footnote{A detailed description of the system
can be found at the AstroWISE portal: www.astro-wise.org/portal}
\citep[AstroWISE,][]{astrowise}. A short description of the main steps
of the data reduction process is given here.

\begin{itemize}
\item Bias subtraction:\\
For each night a master bias image was created from between 10 and 42
raw bias frames after doing an overscan correction and outlier
rejection. These master bias frames were subtracted from all
overscan-corrected science frames to remove 2-D bias patterns.
\item Flat-fielding:\\
Master dome and twilight flats were constructed for each night from
between 5 to 20 exposures. The master dome and twilight flat-fields
are combined into a master flat-field used to correct the science
frames.
\item Astrometry: \\
The astrometry was derived using routines from the Leiden Data
Analysis Center (LDAC). A preliminary astrometry is first estimated as
a simple shift of the 8 chips together to align the field with the
coordinates in the USNO A2.0 catalogue.  The final astrometry for each
chip is derived by using 2nd order polynomials to fit the position of
stars detected in all exposures to their USNO coordinates.
\item Photometry:\\
Since the photometry of the images is scaled to a reference image
during the difference imaging procedure, only mean zero points were
computed in AstroWISE to convert the ADU flux of variable stars to
magnitudes. The zero points were derived as the median of the zero
points obtained from the Landolt standard fields observed during the
whole period of the observations assuming a mean extinction in $R_c$
of 0.09 mag. On average, 4 standard fields were observed per night,
with an average of 100 standard stars landing on each chip. The
zero points and their uncertainties are listed in Table
\ref{tab:zero_points}. The uncertainties are the standard deviations
of the zero points derived from all 40 standard field observations.

\begin{table}
\begin{center}
\caption{Mean zero points and their uncertainties}
\begin{tabular}{l c c}
\hline
\hline
Chip & Zero Point  &  Uncertainty \\
~ & (mag) & (mag) \\
\hline
ccd50 &   24.238   &  0.040\\
ccd51 &   24.334   &  0.048 \\
ccd52 &   24.113   &  0.040 \\
ccd53 &   24.312   &  0.035 \\
ccd54 &   24.311   &  0.029\\
ccd55 &   24.363   &  0.028\\
ccd56 &   24.293   &  0.034\\
ccd57 &   24.171   &  0.029 \\
\hline
\end{tabular}
\label{tab:zero_points}
\end{center}
\end{table}

\item Resampling:\\
Finally, images were regridded to a pixel scale equal to
$0.2^{\prime\prime}$\mbox{$\mathrm{pix^{-1}}$} using the {\it
LANCZOS3} algorithm in SWARP\footnote{For information on SWARP see
  http://terapix.iap.fr/}.
\end{itemize}

\subsection{Difference imaging photometry}

\begin{figure*}
\centering
\includegraphics[width=14cm]{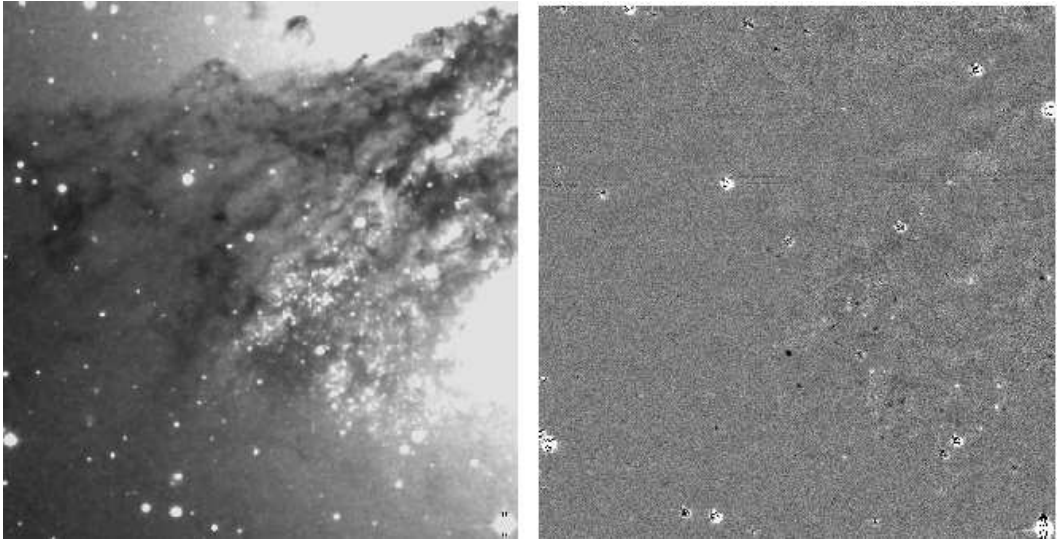}
\caption{Example of an area in Centaurus A and its difference
  image. {\it Left:} detail of the reference image showing the dust
  lanes and star forming regions to the south-east of the centre of
  Centaurus A. {\it Right:} detail of the difference image of June 8th
  2005 showing the same region of Centaurus A.  }
\label{fig:cutouts}
\end{figure*}

For the difference image analysis the same pipeline is used as was
employed by \cite{dejong06} in their analysis of M 31 data. This method
is based on the procedures for matching the point spread function
(PSF) between two images described by
\cite{tomaneycrotts} and although it has a different approach to the
problem of PSF-matching of two images from the method developed by
\cite{alard00} it produces difference images of similar quality.  The
difference image pipeline runs in IRAF and makes use of the DIFIMPHOT
package written primarily by Austin Tomaney. Taking the reduced images
as input, the pipeline goes through the following steps.\\
\begin{itemize}
\item The images are aligned to a common astrometric reference frame
to an rms precision better than 0.1 pixels using stars spread over the
whole image. \\
\item Using cuts on the seeing, astrometric shift and
sky brightness, the best $\sim$15 exposures are selected and combined
into a high signal-to-noise reference image.\\
\item All images obtained
during the same night are combined into one image. These nightly image
stacks will provide the photometric measurements for the detection of
variable sources and their light-curves. Thus, since we use data
obtained during 10 nights we have 10 temporal {\it epochs} in the
subsequent analysis.\\
\item After photometric calibration and matching of the
PSF the reference image is subtracted from each nightly stack,
resulting in a difference image for each night. The PSF of the nightly
stacks and the reference image is measured from bright but unsaturated
stars on the image. From these PSFs a convolution kernel is derived in
Fourier space to convert the better seeing image to the seeing of the
worst seeing image before subtraction. To cope with PSF variations
across the chips, each chip is divided in 16 subregions, for each of
which the PSF matching is done separately, using a minimum of 10
stars. Photometric calibration is done using a subset of non-variable
bright stars. The resulting difference images are dominated by shot
noise in which variable sources show up as positive or negative
residuals.\\
\item Variable sources are detected on the difference images using
SExtractor \citep{sextractor}, where we select groups of 3 adjacent
pixels that are at least 3$\sigma$ above or below the local background. Bright
stars, saturation spikes and bad pixels are masked out during this step.\\
\item Using PSF-fitting photometry the brightness of all variable
sources is measured from the difference images and light-curves are
constructed.
\end{itemize}

\section{Results}
\label{sec:results}

\subsection{Difference image quality}
\label{subsec:dimquality}

\begin{figure}
\centering
\includegraphics[width=8cm]{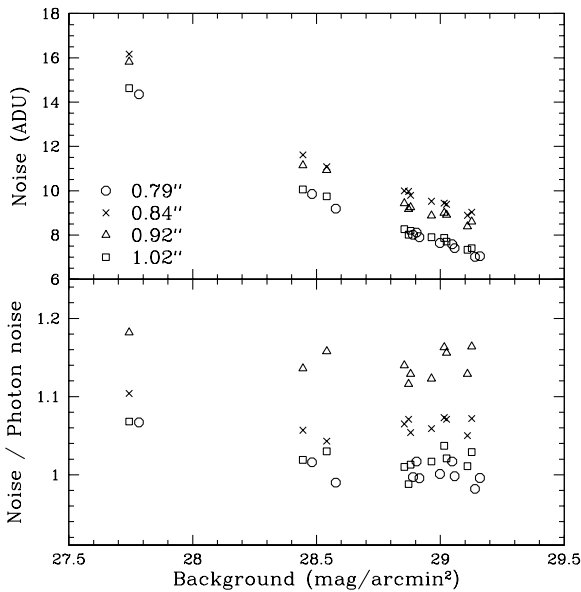}
\caption{Difference image noise levels for a set of regions with
  different background intensity. {\it Top: } standard deviation of
  pixel values for four epochs with different seeing values: May 11
  2005 (0.84\arcsec, crosses), May 12 2005 (0.92\arcsec, triangles),
  June 1 2005 (1.02\arcsec, squares), and June 8 2005 (0.79\arcsec,
  circles). {\it Bottom: } Same as in top panel, but scaled to the
  theoretical photon noise.
}
\label{fig:bgnoise}
\end{figure}

\begin{figure}
\centering
\includegraphics[width=8cm]{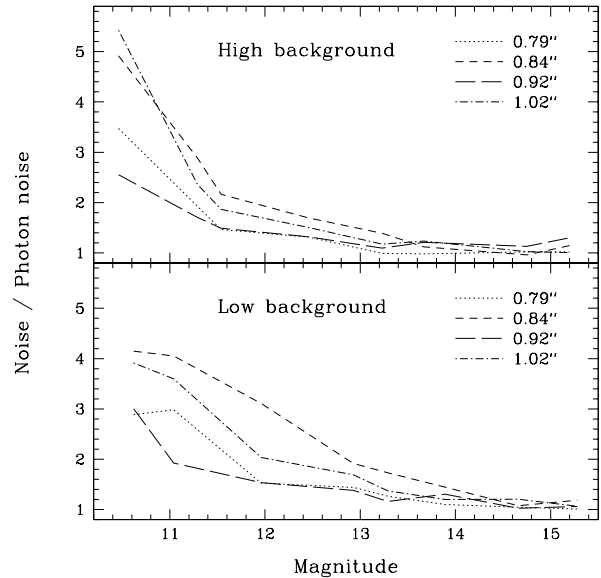}
\caption{Residual strength, in terms of theoretical photon noise, of a
  set of bright, resolved stars of different magnitudes for four
  observation epochs. {\it Top: } Residuals in a high surface
  brightness region ($\mu_R \simeq$28 mag/arcmin$^2$) in the
  difference images of May 11 2005 (short dashed), May 12 2005 (long
  dashed), June 1 2005 (dot-dashed), and June 8 2005 (dotted).  {\it
  Bottom: } Same as top panel, but for a low surface brightness region
  ($\mu_R \simeq$29 mag/arcmin$^2$).  }
\label{fig:starnoise}
\end{figure}

The difference images that result from the above described procedure
are of very high quality, with noise levels close to the theoretical
photon noise.  In Figure \ref{fig:cutouts} a cutout from the reference
image and the corresponding cutout from the difference image of June
8th 2005 are shown. 
While most of the difference image shows only noise, variable
stars in the star forming regions leave positive and negative
residuals. Bright foreground stars also leave residuals in the
difference image due to imperfect subtraction.

As shown theoretically by \cite{gould96}, it is in principle, possible
to create difference images where the noise is dominated by photon
noise. Noise due to systematic effects such as PSF variation and
geometric and photometric alignment can, when handled carefully, be
brought down to levels where they cannot compete with the photon
noise. To assess how well our difference images approach the limit of
pure photon noise, we have determined the noise in several regions of
highly varying background intensity where no obvious variable sources
are present. In the upper panel of Figure \ref{fig:bgnoise} these
noise levels are plotted versus the background intensity for four
different images with different FWHM. There is a strong correllation
with higher noise in brighter regions, which is a clear signature of
the photon noise. Apart from this, there are offsets between the
different epochs, which depend on a combined accuracy of the
PSF-matching, photometric and astrometric alignment, seeing variations
between individual exposures etc.

The lower panel of Figure \ref{fig:bgnoise} shows the same, but now
divided by the theoretical photon noise. For a single exposure, the
photon noise is simply the square root of the number of counts in a
pixel; in this case both the reference image and the nightly images are
stacks of several exposures, so that in each the photon noise should
be the square root of the counts divided by the square root of the
number of exposures going into the stack.
The quality of our difference images turns out to be very good, with
the noise very close to the theoretical photon noise. There might be a
slight degradation at very high background levels, because very close
to the centre of Centaurus A the number of stars that can be used to
construct an accurate PSF model is low, due to saturation of the chips.

Another interesting test is to look at the residuals around resolved,
bright, but unsaturated stars. Because of their higher brightness,
secondary effects such as imperfections in the PSF-matching kernel or
photometric scaling will show up more strongly. Figure
\ref{fig:starnoise} shows the noise divided by the theoretical photon
noise for stars of varying brightness in a high background region
($\mu_R \sim 28.0$ mag/arcmin$^2$, upper panel) and in a low
background region ($\mu_R \sim 29.0$ mag/arcmin$^2$, lower panel). In
both panels the residuals are much higher than the photon noise for
bright stars, and decreasing to the photon noise level at faint
magnitudes. In the high background region this decrease is faster
because the photon noise is higher. That the residuals at the bright
end are higher in the high background region is caused by the poorer
PSF modelling near the centre of Centaurus A, due to the lack of
sufficiently bright, unsaturated stars.  It should be noted that this
problem does not affect image subtraction methods based on direct
pixel-to-pixel fits, such as the one developed by \cite{alardlupton}
and \cite{alard00}.

\subsection{Variable sources}
\label{subsec:variables}

\begin{figure*}
\centering
\includegraphics[width=12cm]{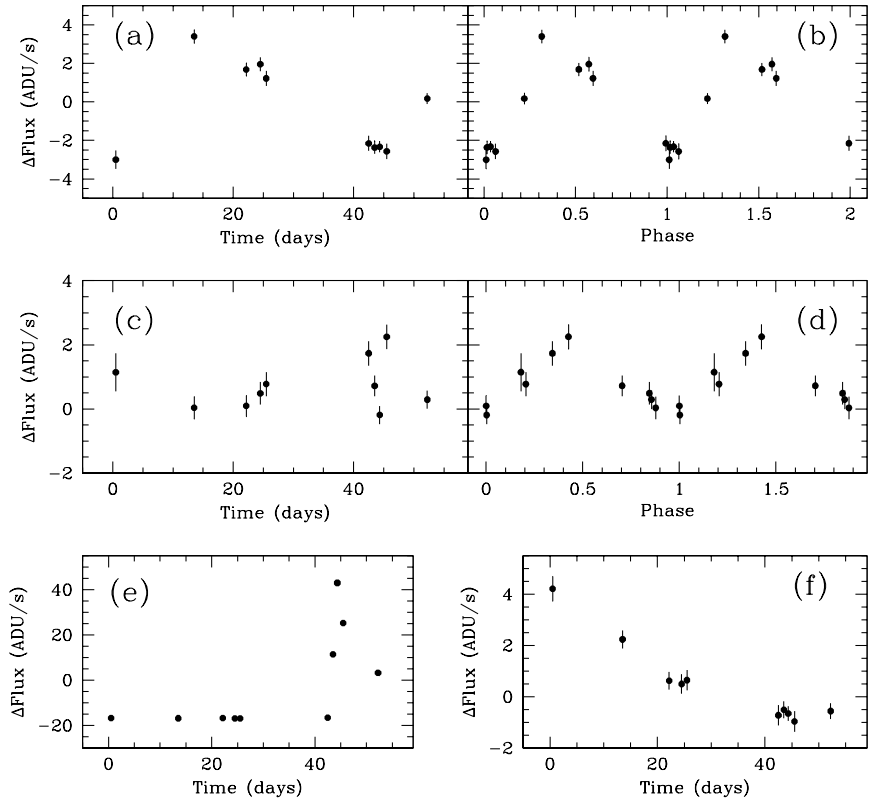}
\caption{Selection of light-curves from variable stars in Centaurus
  A, showing difference flux in ADU/s as a function of the time in
  days from the start of the observations on April 17th 2005. {\it(a)}
  Bright Cepheid-like variable with a period of 43 days. {\it(b)} Same
  variable as in (a) but showing two periods of the folded
  light-curve. 
  {\it(c)} Cepheid-like variable with a period of 2.8
  days. {\it(d)} Same variable as in (c) but showing two periods of
  the folded light-curve. {\it(e)} Nova variable, erupting on May 29
  and reaching its maximum around May 31 2005. {\it(f)} Long-period variable
  reaching its minimum at the beginning of June 2005, with the maximum
  occurring outside our time range. }
\label{fig:lightcurves}
\end{figure*}

Residuals were detected in the difference images using Sextractor and
at all positions with at least two detections light-curves were built. A
by-eye inspection of light-curves and difference images to reject
spurious detections due to bad pixels or foreground stars, resulted in
a catalogue of 271 variable sources. A catalogue with the positions of the
variables is published in electronic form with this article.

The majority of these variables
are expected to be Cepheids and long period variables (LPVs) such as
Mira variables, since these classes are both common and bright. There
is also one Nova in the sample. That we would catch one Nova during
its peak is not unexpected considering the Nova rate of $\sim$28 per
year for Centaurus A \citep{ciardullo90}. With ten points, the
sampling of the light-curves is too sparse to allow an accurate period
and amplitude determination for the Cepheids. For the LPVs the
situation is worse since their periods are longer than the total time
span of our observations. We did, however, apply two different period
finding algorithms to the light-curves, namely the multi-harmonic
periodogram method described by \cite{czerny} and the normalised
periodogram method developed by \cite{lomb}, both of which are
designed to work on unevenly sampled data.  For 29 variables the
periods determined with the two methods lie within 10\% of each other
and between 2 and 50 days. These limits exclude periods that are
longer than the observing period and the aliases that occur close to
one day since this is the minimum spacing between the data
points. 
These reasonably robust periods, which are in the range of typical Cepheid
periods, 
are listed in the electronic catalogue of
variables. Of the other 242 variables, the majority of the ones in the
star forming regions seem chaotic. These are most likely periodic
variables for which the periods cannot be determined because of poor
phase coverage. The majority of the variables away from the centre are
monotonically increasing or decreasing and can be classified as LPVs.

\begin{figure}
\centering
\includegraphics[width=8cm]{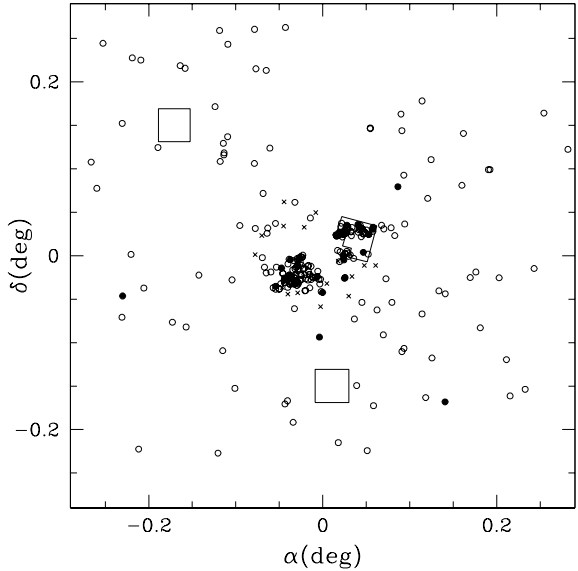}
\caption{Positions of the 271 variable sources detected in our survey,
  with respect to the centre of Centaurus A. Closed circles indicate
  variable sources for which a reasonably accurate period could be
  determined. The two squares outline the fields in which
  \cite{rejkuba03} detected a total of 1504 long-period variables
  using ISAAC at the VLT; the WFPC2 field-of-view used by
  \cite{ferrarese07} in their survey for Cepheids is located just to
  the right from the centre of the Figure. Crosses indicate the
  positions of Nova variables taken from \cite{artyukhina96}.  }
\label{fig:positions}
\end{figure}

\begin{figure}
\centering
\includegraphics[width=8cm]{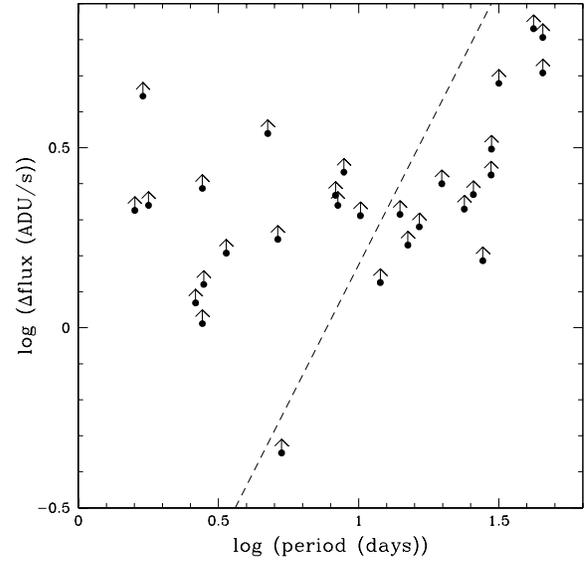}
\caption{Periods and amplitude estimates of the 29 variables
  with robust period determinations. The dashed line represents the
  approximate period amplitude relation found for type I Cepheids in M
  31 \citep{dejong05}, scaled to the distance of Centaurus A and
  corrected for the different gains.}
\label{fig:perampl}
\end{figure}

In Figure \ref{fig:positions} the positions of all 271 variable
sources are plotted. There is a clear concentration of variables in
the star forming regions in the dust ring in the central parts of
Centaurus A. The majority of the stars with confirmed periods between
2 and 50 days are located in these regions. This distribution is not
unexpected, since Cepheids are massive stars with masses of at least 3
or 4 times that of the Sun and up to several tens of solar masses, and
therefore quite young. On the other hand, LPVs belong to old stellar
populations and are expected to be more evenly distributed. Also shown
in Figure \ref{fig:positions} are the locations of the fields
monitored for variability by \cite{rejkuba03} and \cite{ferrarese07}.
Using ISAAC at the Very Large Telescope (VLT), \cite{rejkuba03} were
able to detect 1504 LPVs in an area just over 10 square
arcminutes. None of these are detected in our survey, due to the lower
sensitivity of our data and the fact that LPVs are significantly
fainter at optical wavelengths compared to the
infrared. \cite{ferrarese07} used the Wide Field and Planetary Camera
2 (WFPC2) on board the Hubble Space Telescope to detect Cepheid
variables, resulting in a sample of 56 Cepheids and 70 other periodic
variables. Due to the difference in sensitivity and the fact that
\cite{ferrarese07} use clear periodicity as one of their detection
criteria, we only identify three stars that are both in our and in
their sample. For completeness, also the Nova variables from the
extragalactic part of the General Catalogue of Variable Stars
\citep{artyukhina96} are plotted.  To give the reader an impression of
the quality of the light-curves, we present the light-curves of two
periodic variables, a Nova and an LPV in Figure
\ref{fig:lightcurves}. The Nova variable shown in panel (e) of the
figure has a peak difference flux corresponding to $m_R$=20.2, making
this by far the brightest variable in our sample.  A period-amplitude
plot is shown in Figure \ref{fig:perampl} for the 29 variables with
robust periods.  The amplitudes were determined by taking the
difference between the highest and the lowest point of the lightcurve;
these amplitudes estimates are lower limits since the actual maxima
and minima might not be sampled by the data.  The dashed line
indicates the period-amplitude relation found for type I Cepheids in M
31 using the same techniques as employed here \citep{dejong05}.

\subsection{Detection efficiency}
\label{subsec:deteff}

To determine the detection efficiency for variable sources we put
artificial residuals in the difference images and try to ``detect''
them. Of course, there are different ways to detect sources in
variability studies such as pixel lensing surveys. The method used in
this paper to detect variable stars searches the difference images for
groups of pixels that exceed a certain threshold above or below the
background. This method was also used by the MEGA collaboration in
their M 31 microlensing survey \citep[cf.][]{dejong06}. In another
frequently used method, light-curves are built at each pixel position in
the image, after which the light-curves are identified that show a variable
signal \citep[cf.][]{calchi05}.  To be as general as possible we will
use both approaches in our detection efficiency analysis and show that
the difference in sensitivity between the two is small.

We select a subset of six difference images for the detection
efficiency tests that are of high quality and from CCDs in the centre
of the array. The images are from the nights of April 30, May 30 and
June 8 2005 and correspond to chips 2 and 7 from the CCD array. 
This way reductions in sensitivity due to poor seeing and PSF
degradation in the perifery of the focal plane are avoided.
Although these kind of effects are unavoidable in real surveys, the
detection efficiencies we retrieve will represent upper limits for the
current observational setup.
Because the background from unresolved stars varies strongly in
intensity across the field, the detection efficiency will vary
spatially. To probe this effect we divide both chips in the part where
the background surface brightness is lower than $\mu_R$=28.5
mag/arcmin$^2$ and the part where it is higher, and consider these
regions separately. The high background region includes the small but
very bright centre of Centaurus A. We randomly place residuals of a
fixed (difference) flux in the selected difference images, 1\,000 at a
time in the low background regions and 250 at a time in the smaller
high background regions. Fluxes probed are 2, 1.5, 1, 0.75, 0.5, and
0.25 ADU/s, corresponding to 23.6, 24.0, 24.4, 24.7, 25.2 and 25.9
magnitudes; in total we put 20\,000 residuals with these fluxes in
each region, respectively.
The actual PSFs measured from the stacked nightly
images, which are divided in subregions to deal with possible PSF
variations across the field, are used for building the residuals.
Photon noise is included in the residuals.

After inserting the residuals, SExtractor is run on the difference
images and groups of 3 adjacent pixels that are at least 3$\sigma$
above the local background are selected. Comparison of the list of
detected sources with the input list of residuals gives the detection
efficiency of this residual detection method. At each position where
an artificial residual is inserted, we also perform PSF-fitting
photometry to measure the flux at that position. We consider the
residual to be detected if the recovered flux is at least 3 $\sigma$
above zero, where the $\sigma$ is the photometry error. This error is
calculated by the PSF-fitting photometry routine and based on the
statistics of the fit and includes a correction for photometric
scaling inaccuracies during the image subtraction. The fraction of
sources detected with this PSF-fitting photometry gives a second
measurement of the detection efficiency.

\begin{figure}
\centering
\includegraphics[width=7.5cm]{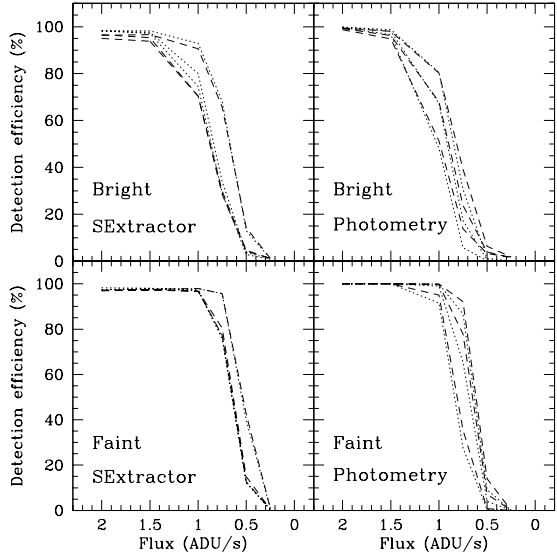}
\caption{Detection efficiencies in percent as function of difference
  flux in ADU/s. Dashed lines show the results for chip 7 in the
  WFI array, dotted lines those for chip 2, where each line
  corresponds to a different epoch. The left panels show the
  efficiencies for the case where residuals are detected as groups of
  significant pixels in the difference images, the right panels for the
  case of PSF-fitting photometry. In the upper panels the results for
  the regions with a background surface brightness higher than 28.5
  mag/arcmin$^2$ are plotted, in the lower panels for the regions with
  lower surface brightness.
}
\label{fig:deteff}
\end{figure}

Figure \ref{fig:deteff} shows the detection efficiencies for both
detection methods and all regions.
There is indeed a difference in the detection efficiency in the bright
regions (upper panels) and the fainter regions (lower panels),
although not very large. In both cases the detection efficiency is
100\% down to a flux of 1.5 ADU/s and virtually 0\% at 0.5 ADU/s, with
the transition more gradual in the bright regions. This more gradual
slope is due to the larger variation in the background
intensity. These results show that even very close to the bright
centre of Centaurus A, residuals with a flux of 1.5 -- 1 ADU/s are
still detected with high efficiency. There is not much difference between the
SExtractor-based detection method (left panels) and the
photometry-based method (right panels).
A variability survey with the WFI at the ESO/MPG 2.2m telescope with
9\,000$s$ exposure time per epoch thus will have a detection limit of
at best slightly below 1 ADU/s. This corresponds to $\sim$24.5 $m_R$,
or for a variable of absolute magnitude $M_R$=$-4$ to a magnitude
difference in the order of 1 magnitude. Since the majority of Cepheids
has amplitudes of less than 1 magnitude this means that only the very
bright ones or ones with high amplitudes are detected.

\section{Implications for microlensing surveys}
\label{sec:microlensing}

Now that we have shown that Difference Image Photometry works well at
the distance of Centaurus A, we want to investigate the prospects of
ground-based microlensing surveys towards this galaxy. In the previous
section the detection efficiency for a single residual of a certain
flux in a high quality difference image was determined. Taking this as
the detection efficiency for a microlensing event of the same maximum
amplitude is of course a gross oversimplification. To be able to
``detect'' a microlensing event, enough information must be obtained
from the light-curve to be reasonably certain of the lensing nature of
the detected variability. In practice one would want to detect the
event in more than one difference image and with sufficient
signal-to-noise to recover certain characteristic signatures, such as
light-curve shape and achromaticity. Even assuming a survey with
observations every night, a detection limit of 1 ADU/s is optimistic,
but can serve as an upper limit.

In the light of the arguments discussed in Section \ref{sec:intro} we
concentrate on microlensing by objects in the halo of Centaurus A.
If we assume the halo is a cored isothermal sphere
with velocity dispersion $\sigma$ and core radius $r_c$, the density
is given by,
\begin{equation}
\rho(r)= \frac{\sigma^2}{2 \pi G} ~ \frac{1}{r^2+r_c^2}.
\label{eq:rhohalo}
\end{equation}
By definition, a source is considered to be microlensed if it is lying
within the Einstein radius of a lens, i.e. if it is magnified at least
a factor of 1.34.  The microlensing optical depth $\tau$ is defined as
the number of lenses within one Einstein radius of the line-of-sight
towards the source and therefore gives the fraction of stars that is
microlensed at a given time. This means that $\tau$ is given by,
\begin{equation}
\tau = \int_{0}^{r_h} n_l(r) \pi R_{\rm E}^2 ~dr ~=~ \int_{0}^{r_h}
\frac{f_b\rho(r)}{M_{\rm lens}} \frac{4 \pi G M_{\rm lens}}{c^2} \frac{D_{\rm LS}D_{\rm OL}}{D_{\rm OS}} ~dr,
\end{equation}
where $n_l$ is the number density of lenses, $f_b$ is the fraction of
halo mass in the lenses, $R_{\rm E}$ is the Einstein radius, and $D_{\rm OL}$,
$D_{\rm OS}$ and $D_{\rm LS}$ are the distances between observer, lens and
source. We integrate from the position of the source to the outer
radius of the halo $r_h$.  For external galaxies like Centaurus A,
$D_{\rm OL} \simeq D_{\rm OS}$ and $D_{\rm LS} \simeq r$, so that,
\begin{equation}
\tau(r_0) \sim \frac{2 f_b \sigma^2}{c^2} \int_0^{\sqrt{r_h^2-r_0^2}}
\frac{r}{r^2 + r_0^2 + r_c^2} ~dr,
\end{equation}
for a source star located at a projected distance $r_0$ from the
centre of the galaxy.
Integrating gives,
\begin{equation}
\tau(r_0) ~=~ \frac{f_b \sigma^2}{c^2} ~ln \left( \frac{r_h^2 +
  r_c^2}{r_0^2 + r_c^2} \right).
\end{equation}
For a MACHO mass fraction $f_b$ of 10\%, a halo velocity dispersion
of $\sim$100 km/s \citep{peng04} and reasonable values for $r_h$,
$r_0$ and $r_c$ this gives values in the order of $10^{-7}$, meaning
that at any time one in ten million stars is being microlensed.

However, at the distance of Centaurus A we are in the regime where
only very high amplification events are detected, the so-called spike
pixel-lensing regime defined by \cite{gould96}.  In this regime the
impact parameter $\beta$, expressed in units of the Einstein Radius
$R_{\rm E}$, is very small, $\beta \ll 1$, and inversely proportional to the
maximum amplification $A$: $\beta \sim 1/A$.
Suppose a source has to change in brightness by a certain amount
$L_{\rm det}$ in order to be detected. This means that for a microlensing
event to be detected, the maximum impact parameter is $\beta \sim 1/A
~=~ L/L_{\rm det}$ for a source star with luminosity $L$.
The number of detectable microlensing events at any time is now given
by the optical depth multiplied by the square of the impact parameter,
\begin{equation}
N_{\rm det} \simeq \tau \cdot \beta^2(L) = \tau \int dL f(L) \frac{L^2}{L_{\rm det}^2},
\end{equation}
where $f(L)$ is the luminosity function of the source stars.
Using the fact that the luminosity of the galaxy $L_{gal}=\int dL f(L)
L$ and the fluctuation brightness of the stellar population
$\bar{L}=\int dL f(L) L^2/L_{gal}$ we obtain,
\begin{equation}
N_{\rm det} \simeq \tau \frac{\bar{L} L_{gal}}{L^2_{\rm det}}.
\end{equation}
The absolute magnitude of Centaurus A is $M_R \simeq -22$
\citep{laubertsvalentijn} and a typical value for the fluctuation
brightness of an old stellar population is in the order of $R \simeq
-1.3$ \citep[e.g.][]{buzzoni93,tonry01}. 
Given our detection limit of $m_R\simeq$24.5, corresponding to
$M_R\simeq$-3.5 this leads to a number of detectable microlensing
events at any time of $N_{\rm det} = 10^{6.5} \tau$. Assuming a MACHO
halo mass fraction of about 10\% this means that at any moment $\sim$0.3
microlensing events can be seen.

The rate $\Gamma$ at which microlensing events due to halo lensing
occur is given by the number of events $N_{\rm det}$ divided by the average
duration of the events, measured by the time during which the
magnification is above half the maximum magnification $t_{\rm FWHM}$. In
the regime where $\beta \ll 1$, we can approximate $t_{\rm FWHM} =
\sqrt{3} \beta t_{\rm E}$ \citep{gondolo99}. Here $t_{\rm E}$ is the Einstein
crossing time, the time it takes the lens to cross the full Einstein
disk, given by,
\begin{eqnarray}
t_{\rm E} ~=~ \frac{2 R_{\rm E}}{v_{\perp}} ~=~ \frac{2}{v_{\perp}}
\sqrt{ \frac{4 G M_{\rm lens}}{c^2} \frac{D_{\rm LS}D_{\rm OL}}{D_{\rm OS}}}\\
\simeq 2.8 \times 10^7 s ~ \left( \frac {M}{M_\odot} \right)^{1/2} ~ \left(
\frac {D_{\rm LS}}{10 kpc} \right)^{1/2} ~ \left( \frac {v_\perp}{100 km/s} \right)^{-1},
\end{eqnarray}
where $v_\perp$ is the transverse velocity of the lens with respect to
the observer-source line of sight.
A typical value of $t_{\rm E}$ for 1 $M_{\odot}$ halo objects in Centaurus A
is therefore in the order of a year.
The event rate is now given by,
\begin{eqnarray}
\Gamma \simeq \tau \int dL f(L) \beta^2 / t_{\rm FWHM} \simeq
\sqrt{3}\frac{\tau}{t_{\rm E}} \int dL f(L) \frac{L}{L_{\rm det}}
\nonumber\\
~\simeq~ \sqrt{3}\frac{\tau}{t_{\rm E}} \frac{L_{\rm gal}}{L_{\rm det}}
\end{eqnarray},
and filling in the numbers gives $\Gamma \simeq \sqrt{3}\cdot10^{7.4} \tau /
t_{\rm E}$. Again assuming an $f_b$ of 10\% we get an expected detectable
event rate of $\sim$4 per year. The average event duration is simply
$N_{\rm det}/\Gamma \sim 1$ month.

 An added complication for studying microlensing due to MACHOs in
the halo of Centaurus A is that these microlensing events have to be
separated from different possible contaminants, such as variable stars
and lensing by other lens populations. Most periodic variables can be
recognised as such by extending the time baseline of the survey
sufficiently. Cataclysmic variables are not period, but can be
identified based on the lightcurve shape and colour
evolution. Considering the expected Nova rate for Centaurus A of
$\sim$28 per year \citep{ciardullo90}, which is several factors higher
than the halo microlensing rate, the need for sufficiently dense time
sampling and multi-band photometry is clear.
Microlensing by stars in Centaurus A, rather than by objects in the
halo, will also have to be disentangled from the halo lensing signal.
As for M 31 and M 87, this so-called ``self-lensing'' will dominate
over halo lensing in the central parts of Centaurus A
\citep[cf.][]{baltz03,baltz04,riffeser06}. The cleanest halo lensing
sample is therefore obtained in the outer parts of the galaxy.

\section{Conclusions}
\label{sec:conclusions}

In this paper we have demonstrated that difference imaging techniques
can be successfully applied to ground-based data of Centaurus A, a
giant elliptical galaxy at a distance of 4 Mpc. Using the Wide Field
Imager at the ESO/MPG 2.2m telescope at La Silla, and a total exposure
time of 9\,000$s$ per epoch, we reach a difference flux detection
limit corresponding to a brightness of $m_R\simeq$24.5.  We have
detected and photometered 271 variable sources in Centaurus A, mostly
Cepheids and long-period variables and a Nova. The majority of
these variables were previously unknown and are available as an online
catalogue with this paper. Clearly, these techniques can be used as
powerful tools to enlarge the samples of known variable stars in and
to study stellar populations of Centaurus A and other galaxies in the
nearby universe.  At these depths the difference images do not suffer
from crowding, and can be photometered straightforwardly.

The microlensing optical depth towards Centaurus A due to lenses in
the halo is of the order of $\tau \sim 10^{-7}$ if we assume that 10\%
of the halo mass is in compact objects. Given this optical depth and
our detection limit the expected microlensing event rate is $\sim$4
per year over the whole galaxy. At large distances it will be
impossible to identify the source stars of individual microlensing
events and blending by stars and variables that are too faint to be
detected will play a more important role. Also, only very high
amplification events can be observed, meaning that the microlensing is
completely in the spike regime \citep{gould96}. Even more than for
microlensing in M 31, the information that can be obtained for
individual events is limited. Statistical analysis of event rates,
spatial distribution and timescale distribution is the only way of
extracting scientific information out of these data. This again
emphasises the need for large samples of microlensing events.

The pilot experiment described here shows that it would be possible to
put useful limits on the number of microlensing events in Centaurus A,
but a survey that is several magnitudes deeper would be required. This
could be achieved with a large ground-based telescope equipped with a
wide-field imager large enough to take in all of the galaxy in one
exposure, or better still, with a wide-field space observatory.
We therefore endorse the plea of \cite{baltz05} for inclusion of
microlensing of nearby galaxies into the observing programmes of
upcoming wide-field cosmological survey facilities.

\begin{acknowledgements}
We thank the anonymous referee for useful comments and suggestions to
improve this paper. 
J.T.A.d.J. acknowledges support from DFG Priority Program 1177.
Ph.H\'eraudeau acknowledges the financial
support provided through the European Community's Human Potential
Programme under contract HPRM-CT-2002-00316, SISCO.
\end{acknowledgements}

\end{document}